\documentclass{elsart}
\usepackage{amssymb}
\usepackage{amsmath}
\newcommand{\be}{\begin{equation}}
\newcommand{\ee}{\end{equation}}
\newcommand{\bea}{\begin{eqnarray}}
\newcommand{\nn}{\nonumber}
\newcommand{\eea}{\end{eqnarray}}

\begin{document}

\begin{frontmatter}

% Title, authors and addresses

% use the thanksref command within \title, \author or \address for footnotes;
% use the corauthref command within \author for corresponding author footnotes;
% use the ead command for the email address,
% and the form \ead[url] for the home page:
\title{The viability of theories with matter coupled to the Ricci scalar}
% \thanks[label1]{}
\author{Thomas P.~Sotiriou}
\ead{sotiriou@umd.edu}
% \ead[url]{home page}
% \thanks[label2]{}
% \corauth[cor1]{}
\address{Center for fundamental Physics, University of Maryland, College Park, MD 20742-4111, USA}
% \thanks[label3]{}

%\title{}

% use optional labels to link authors explicitly to addresses:
% \author[label1,label2]{}
% \address[label1]{}
% \address[label2]{}

%\author{}

%\address{}

\begin{abstract}
% Text of abstract
Recently there has been a proposal for modified gravitational   $f(R)$ actions which include a direct coupling between the matter action and the Ricci scalar, $R$. Of particular interest is the specific case where both the action and the coupling are linear in $R$. It is shown that such an action leads to a theory of gravity which includes higher order derivatives of the matter fields without introducing more dynamics in the gravity sector and, therefore, cannot be a viable theory for gravitation. 
\end{abstract}

\end{frontmatter}

Even thought alternative theories of gravity have always been a subject of study since the introduction of General Relativity (GR), nowadays they appear to attract increased interest. This is mainly due to the fact that they appear as candidates for explaining the nature of dark matter and dark energy and, consequently, they might bring us closer to answering some of the most important open questions of late  time cosmology. 

One of the proposals which have been given much attention recently is that of $f(R)$ theories of gravity: theories described by the action
\be
\label{faction}
S_f=\int d^4 x \sqrt{-g}[\frac{1}{2\kappa} f(R) + L_m(g_{\mu\nu},\psi)]
\ee
where $g$ is the determinant of the metric $g_{\mu\nu}$, $\kappa=8\pi\, G$, $f(R)$ is a general function of the the Ricci scalar $R$, $L_m$ is the matter action and $\psi$ collectively denotes the matter fields. There exist more than one variational principle which could be applied to the above action. Namely, one could use the standard metric variation, in which case the result is the so-called metric $f(R)$ theories of gravity, or an independent variation of the metric and the affine connection (Palatini variation) which leads to Palatini $f(R)$ gravity
 (see \cite{buchdahl,ffv,Nojiri:2006ri,Capozziello:2007ec,Sotiriou:2007yd} for early works and reviews\footnote{It is worth pointing out that $f(R)$ gravity in the metric formalism appears to be unique among higher order gravity theories in that it avoids the so-called Ostrogradski instability \cite{woodard}.}). In the Palatini formalism the connection $\Gamma^\lambda_{\phantom{a}\mu\nu}$ is independent from the metric. Therefore, $R$ is replaced in action (\ref{faction}) by
 ${\mathcal R}=g^{\mu\nu} {\mathcal R}_{\mu\nu}$, where ${\mathcal R}_{\mu\nu}$
   is the Ricci tensor constructed  with the independent connection. Note that, in action (\ref{faction}), $L_m$ is taken {\em a priori} to be independent of the connection. If this assumption is abandoned the outcome is yet another version of $f(R)$ theory, metric-affine $f(R)$ gravity \cite{Sotiriou:2006qn}. Needless to say that all three version of $f(R)$ gravity are different theories, despite the similarity of their actions \cite{Sotiriou:2006hs,Sotiriou:2006sr}.
   
Much work has been done in studying the phenomenology and the viability of $f(R)$ gravity (see \cite{list} for some examples). Even though it is not very likely that some action within this class can constitute a complete and viable alternative to GR, $f(R)$ theories have proved to be very useful as toy-theories of gravity. As such, they can significantly contribute to out understanding of the gravitational interaction.

Recently, the $f(R)$ idea was generalized to include a coupling between the Ricci scalar and the matter itself. The action presented in \cite{Bertolami:2007gv} was
\be
\label{action}
S=\int \sqrt{-g}d^4 x \left\{\frac{1}{2} f_1(R)+\left[1+\lambda f_2(R)\right] L_m\right\},
\ee
where, $f_{1,2}$ are arbitrary functions of the $R$, $L_m$ is the matter action and $\lambda$ acts as a coupling constant. Other actions with similar couplings between $R$ and matter have been studied in \cite{Mukohyama:2003nw,Dolgov:2003fw,Nojiri:2004bi}. 

Variation of the action (\ref{action}) with respect to the metric yields
\begin{align}
\label{field}
F_1(R) R_{\mu\nu}-\frac{1}{2} f_1(R) g_{\mu\nu}-\nabla_\mu\nabla_\nu F_1(R)+g_{\mu\nu}\Box F_1(R)\nn\\
= -2\lambda F_2(R) L_m R_{\mu\nu}+2\lambda (\nabla_\mu\nabla_\mu-g_{\mu\nu}\Box) L_m F_2(R)\nn\\
+[1+\lambda f_2(R)] T_{\mu\nu},
\end{align}
where $F_{1,2}(R)\equiv f'_{1,2}(R)$, the prime denotes differentiation with respect to the argument and 
\be
T_{\mu\nu}\equiv -\frac{2}{\sqrt{-g}}\frac{\delta (\sqrt{-g} L_m)}{\delta g^{\mu\nu}}
\ee
is the matter stress-energy tensor as usual. Note that $T_{\mu\nu}$ is not divergence-free here, due to the direct coupling between matter and $R$. Therefore, the theory will exhibit violations of the Einstein Equivalence Principle \cite{willbook}, which, however, can in principle be controlled by tuning $\lambda$ \cite{Bertolami:2007gv}.

In \cite{Bertolami:2007gv} the authors examined whether theories described by the action (\ref{action}) could account for phenomenology usually attributed to dark matter. In \cite{Bertolami:2007vu} the consequences of such a theory in stellar equilibrium were studied. As a matter of fact, the action used there was a particular case of action (\ref{action}): the one in which\footnote{We are using here the conventional notation for $\kappa$, whereas in \cite{Bertolami:2007gv,Bertolami:2007vu} $\kappa$ is taken to be $(16\pi\,G)^{-1}$.} $f_1(R)=\kappa^{-1} R$ , and $f_2(R)=R$. It was argued in \cite{Bertolami:2007vu} that this linear choice for $f_1$ and $f_2$ should serve as a good approximation, since, firstly it was expected that the phenomenology related to the coupling should overwhelm any phenomenology coming from the non-linearity of $f_1$ and, secondly, Solar System and other equivalence principle related tests will force   $\lambda f_2\ll 1$, which combined with the first requirement yields $f_2(R)=R$ as the simplest possibility.

With these choices for $f_{1,2}$ the action reads
\be
\label{laction}
S=\int \sqrt{-g}d^4 x \left\{ \frac{R}{2\kappa}+\left[1+\lambda R\right] L_m\right\},
\ee
and the corresponding field equations are
\begin{align}
\label{lfield}
R_{\mu\nu}-\frac{1}{2}R g_{\mu\nu}
= &-2\kappa\lambda L_m R_{\mu\nu}+2\kappa\lambda (\nabla_\mu\nabla_\nu-g_{\mu\nu}\Box) L_m \nn\\
&+[1+\lambda R] \kappa T_{\mu\nu},
\end{align}
Let us first perform a straightforward comparison between eq.~(\ref{field}) and eq.~(\ref{lfield}). When $f_1$ is linear $F_1$ is a constant and, therefore, the last two terms in the left hand side (lhs) of eq.~(\ref{field}) are no longer present in eq.~(\ref{lfield}). Additionally, when $f_2$ is linear, $F_2$ is constant as well, so $R$ is no longer present in the second term on the right hand side (rhs) of eq.~(\ref{lfield}).

What is important to notice, however, is that these are the terms which contained derivatives of $R$. Indeed, eq.~(\ref{field}) is fourth-order in the metric, containing second derivatives of $R$. On the contrary, eq.~(\ref{lfield}) is just second order in the metric, exactly like GR. 

Eq.~(\ref{field}) differs from Einstein's equation in another important way as well: it contains higher derivative of the matter fields through the second term in the rhs. This characteristic still remain even when $f_{1,2}$ are chosen to be linear, as can easily be seen in eq.~(\ref{lfield}). Assuming, as usual, that the matter action $L_m$ contains only to first derivatives of the matter fields (so as to lead to second order field equations for the matter fields when varied with respect to them), eq.~(\ref{field}) is a fourth order partial differential equation (PDE) in the metric and a third order PDE in the matter fields. Eq.~(\ref{lfield}) is again a third order PDE in the matter fields but just a second order PDE in the metric. As a reference point, Einstein equations are  second order PDEs in the metric and first order in the matter fields.

A serious issue becomes immediately apparent for the case where $f_{1,2}$ are linear: the differential order in the matter fields is higher than the differential order in the metric! Remarkably this has gone unnoticed until now, even though it is a source of serious problems.

To make this clearer, let us recall some past experience form Palatini $f(R)$ gravity. Applying the Palatini variational principle on action (\ref{faction}) one gets
\begin{align}
\label{field1}
 &f'({\mathcal R})R_{(\mu\nu)}-\frac{1}{2}f({\mathcal R}) g_{\mu\nu}=\kappa T_{\mu\nu}, \\
 &\bar{\nabla}_\lambda\left( \sqrt{-g}f'({\mathcal R})g^{\mu\nu}\right)=0,
\end{align}
where $\bar{\nabla}_\lambda$ is the covariant derivative defined with the independent connection. It is easy to show (see for instance \cite{Sotiriou:2007yd,sing}) that these equations can actually be combined to a single one:
 \begin{align}
\label{eq:field}
{R}_{\mu \nu} -\frac{1}{2}R g_{\mu\nu}\!=&\! \frac{\kappa}{f'}T_{\mu \nu}- \frac{1}{2}g_{\mu \nu}\! 
                       \left(\!{\mathcal R} - \frac{f}{f'} \right)\! +\! \frac{1}{f'} \left(
			\nabla_{\mu} \nabla_{\nu}
			\!- g_{\mu \nu} \Box
		\right)\! f'-\nn\\
& \quad- \frac{3}{2}\frac{1}{f'^2} \left(
			(\nabla_{\mu}f')(\nabla_{\nu}f')
			- \frac{1}{2}g_{\mu \nu} (\nabla f')^2
		\right),
\end{align}
where, however, ${\mathcal R}$ and consequently $f({\mathcal R})$ and $f'({\mathcal R})$ are actually functions of  $T=g^{\mu\nu}T_{\mu\nu}$, due to the fact that the trace of eq.~(\ref{field1})
\be
f'({\mathcal R})R-2f({\mathcal R})=\kappa T,
\ee
relates ${\mathcal R}$ and $T$ algebraically (for instance, for $f({\mathcal R})={\mathcal R}+\epsilon {\mathcal R}^2$, where $\epsilon$ is some parameter with suitable dimensions, $R=-\kappa T$). Therefore, the lhs of eq.~(\ref{eq:field}) depends only on the matter fields and at the same time includes, up to third derivatives, exactly like the lhs of eq.~(\ref{lfield}). At the same time both of these equation are second order PDEs in the metric.

Now, the presence of the higher order derivatives of matter fields in eq.~(\ref{eq:field}) has been shown to lead to very serious shortcomings for Palatini $f(R)$ gravity. The post-Newtonian metric becomes algebraically related to the density \cite{post} making the result of Solar System tests density dependent. This is clearly unacceptable since such test should be valid over a wide range of densities\footnote{In \cite{LMS} concerns have been expressed on whether issues of averaging can have an affect on this but it is still questionable if this is the case.}. At the same time, the presence of these higher order derivatives has been shown to lead to the appearance of singularities on the surface of specific spherically symmetric matter configurations when one attempts to match an interior to the unique exterior solution, making any matching impossible \cite{sing}. This leads to the absence of solution describing the gravitational field of physical objects which can even be described with Newtonian gravity \cite{sing}. Finally, even the problem found earlier in Palatini $f(R)$, the unacceptable modification it appears to introduce in the standard model of particle physics \cite{part}, seems to have its root in the presence of the higher order derivatives of matter fields \cite{sing}.

It has to be stressed that these problems are not specific to Palatini $f(R)$ gravity (or to the dynamically equivalent $\omega_0=-3/2$ Brans-Dicke theory \cite{Sotiriou:2006hs}), but as already mentioned in \cite{sing}, they are generic problem for theories in which the differential order of the matter fields in the field equations is higher than the differential order in the metric. This is because, the unusual behaviour discussed here does not really have to do with the fine details of the field equations, but merely with the fact that this unorthodox differential structure makes gravity non-commulative: the metric ceases to be an integral over the source and becomes algebraically related to the matter fields and their derivatives. Therefore, any discontinuities in the latter, which are in general allowed, will become unacceptable discontinuities and singularities in the metric, leading to unacceptable phenomenology.

Having said these, it is straightforward to realize that the theory described by the action (\ref{laction}) and the field equations (\ref{lfield}) is bound to be burdened with essentially the same problems and no further calculation is actually needed to show that.  Just for the sake of clarity and in order to further verify our claim, let us consider the weak gravity regime.

First of all, notice that if one takes the trace of eq.~(\ref{lfield}) this gives
\be
\label{trace}
R=\frac{6\kappa \lambda \Box L_m -\kappa T}{1+\kappa \lambda (T-2 L_m)}.
\ee
Replacing this back in eq.~(\ref{lfield}) in order to eliminate $R$ yields after some manipulations
\begin{align}
\label{lfield2}
R_{\mu\nu}=\frac{1}{1+2\kappa \lambda L_m}\Bigg[&\kappa T_{\mu\nu} +2 \kappa \lambda \left(\nabla_\mu\nabla_\nu-g_{\mu\nu}\Box\right) L_m+\nn\\
&+\frac{6 \kappa \lambda \Box L_m-\kappa T}{1+\kappa \lambda (T-L_m)}\left(\kappa \lambda T_{\mu\nu}-\frac{1}{2}g_{\mu\nu}\right)\Bigg].
\end{align}
This is, in general, the most usual and convenient form to discussed the weak field regime. Now, 
suppose the metric is taken to be flat plus a perturbation:
\be
g_{\mu\nu}=\eta_{\mu\nu}+h_{\mu\nu}
\ee
We will not attempt here to perform a post-Newtonian expansion. This is a tedious task and not even well established for a theory which exhibits violation of the EEP. Fortunately, this is not needed anyway for our purposes. Keeping just first order terms and 
applying the  standard gauge conditions
\be
\label{gauge1}
h^\mu_{i,\mu}-\frac{1}{2}h^\mu_{\mu,i}=0\;,\qquad
%\label{gauge2}
h^\mu_{0,\mu}-\frac{1}{2}h^\mu_{\mu,0}=\frac12 h^0_{0,0}\;,
\ee
the time-time component of $R_{\mu\nu}$, for instance, reduces to
\be
R_{00}=-\nabla^2{(h_{00})},
\ee
as is well known, where $\nabla^2\equiv\delta_{ij}\partial_i\partial_j$. Focus now on the rhs of eq.~(\ref{lfield2}) and consider for simplicity just the second term in the square brackets. If EEP test are not to show sever violations, $\lambda$ should be chosen in such a way, so as 
\be
\label{cond1}
2\kappa \lambda L_m \ll 1,
\ee
since $(1+2\kappa \lambda L_m)^{-1}$ acts as a coefficient of $T_{\mu\nu}$. Assuming now, as usual, an almost static matter configuration, so that time derivative can be considered of higher order, one straightforwardly sees that the second term in the square brackets  at the rhs of eq.~(\ref{lfield2}) will lead to a term (amongst others) proportional to $\kappa \lambda \nabla^2 L_m$. {\em I.e.}
\be
\label{final}
\nabla^2{(h_{00})}\propto \kappa \lambda \nabla^2 L_m +{\rm E.S.T.},
\ee
where ${\rm E.S.T}$ stand for extra source terms (not necessarily of higher order).

The mere form of eq.~(\ref{final}) is enough to derive conclusions, and this is why a more detailed calculation is not needed: The time-time component of the weak field metric clearly has an algebraic contribution from $L_m$. This implies that the metric is not plainly an integral over the sources, as in GR, but it also has a direct algebraic dependence on the matter field and their derivative (included in $L_m$). Obviously, whether or not Solar System test will be satisfied, or even whether the Newtonian limit itself could be recovered, would have to depend on the special characteristic of the matter, such as the energy density for instance. Additionally, any discontinuities or delta function in the matter could lead to unavoidable singularities! Note also, that tuning $\lambda$ in order to suppress this term cannot really ameliorate things here, since the problem does not really lie on the magnitude of this problematic term, but on its very presence (due to the possibility of exhibiting discontinuities {\em etc.}).

Of course, as already mentioned, this is just one example of the problems that could be caused by the unusual differential structure of eq.~(\ref{lfield}) which we used for demonstrative purposes. It should be clear by now, even just based on the comparison with Palatini $f(R)$ gravity and eq.~(\ref{eq:field}), that the theory described by the action (\ref{laction}) cannot constitute a viable alternative to GR, for essentially the reasons described in Refs.~\cite{sing,post,part} for Palatini $f(R)$ gravity. As it has been already stated in \cite{sing} all of these issues do not constitute a special characteristic of the latter, nor are they specific (or generic) to the Palatini variation. They are just straightforward shortcomings of theories in which gravity does not act in a cumulative way, {\em i.e.~}theories in which the metric is not given as an integral but is also algebraically related to the sources and their derivatives, and therefore is influenced by local properties of the matter. Obviously, all theories in which the differential order of the matter field is higher of that of the metric in the field equations have this characteristic.

Another way to understand this problem is to observe that in action (\ref{laction}) one has essentially added, with respect to GR, a strong coupling between the scalar curvature and matter, without however adding any extra dynamics. This is exactly the case for Palatini $f(R)$ gravity as it becomes clear when the latter is seen as an $\omega_0=-3/2$ Brans--Dicke theory: The (auxiliary) Brans--Dicke scalar essentially introduces a strong coupling between gravity and matter (which, however, does not lead to violation of the equivalence principle) \cite{Sotiriou:2006hs}.

Notice that once any of the function $f_{1,2}$ in action (\ref{action}) are non linear in $R$, the field equations (\ref{field}) immediately become fourth order in the metric. Now, even though the strong coupling still exists, more dynamics are introduced. The field equations are still third order PDEs in the matter fields, but now the differential order in the metric exceed the differential order in the matter fields, and therefore gravity becomes cumulative again. This implies that there is no reason to expect that the issues discussed here will persist. However, a more detailed study of the phenomenology of such theories is needed to arrive to definite conclusions.

Finally, it is also worth pointing out that the action (\ref{laction}) constitutes another example case where the dynamics of theory are not apparent in the action, due to the fact that the chosen representation of the theory is not helpful in this aspect (see \cite{Sotiriou:2007zu} for a general discussion on theories and representations). Indeed, judging for action (\ref{laction}), one could easily be tricked to think that it will lead to a theory which is second order in the metric and only first order in the matter fields (as there are only first derivatives of the latter). However, $R$ includes second order derivatives of the metric. When $R$ is not coupled to other fields (and has no self interactions), in a metric variation, these second derivative terms conveniently lead to a surface term. However, when $R$ is present in coupling terms, by parts integration to arrive to total divergencies introduces extra derivatives of the fields $R$ is coupled to; in this case this is the matter Lagrangian $L_m$.

To summarize, it was argued that a gravitational theory which, in addition to the Einstein-Hilbert term and the matter lagrangian includes also  a direct linear coupling between $R$ and the matter action cannot be viable. This has two important consequences: Firstly, if theories with coupling between $R$ and the matter are to be explored further, either the coupling or the gravitational action have to be taken to be non-linear (with all the complications this might bring). Secondly, a linear action with a linear coupling cannot be used even as an approximation of a more complicated one for specific applications (as, for instance, in \cite{Bertolami:2007vu}), as it leads to spurious phenomenology\footnote{As a matter of fact, in \cite{Bertolami:2007vu}, divergences are found. They are not considered worrying by the authors since they are attributed to the specific equations of state and the limited validity of the perturbative scheme used. However, it seems much more probable that they are of the same nature as the divergences discovered in \cite{sing} for Palatini $f(R)$ gravity.} and is in this sense unique among the general class of theories described by action (\ref{action}) for arbitrary choices of $f_{1,2}$.

{\em Acknowledgements:}
The author would like to thank Valerio Faraoni for enlightening discussions and acknowledge support by the National Science Foundation under grant PHYS-0601800.

%\label{}

% The Appendices part is started with the command \appendix;
% appendix sections are then done as normal sections
% \appendix

% \section{}
% \label{}

\end{document}